\documentclass[fleqn,usenatbib]{mnras}

\usepackage{newtxtext,newtxmath}
\usepackage[T1]{fontenc}

\DeclareRobustCommand{\VAN}[3]{#2}
\let\VANthebibliography\thebibliography
\def\thebibliography{\DeclareRobustCommand{\VAN}[3]{##3}\VANthebibliography}

\usepackage{graphicx}	% Including figure files
\usepackage{amsmath}	% Advanced maths commands
\title[OJ 287]{Absence of the predicted 2022 October outburst of OJ 287 and implications for binary SMBH scenarios}
\author[S. Komossa et al.]{
S. Komossa,$^{1}$\thanks{E-mail: astrokomossa@gmx.de (SK)}
D. Grupe,$^{2}$
A. Kraus,$^{1}$
M.A. Gurwell,$^{3}$
Z. Haiman,$^{4,5}$
F.K. Liu,$^{6,7}$
A. Tchekhovskoy,$^{8}$
\newauthor
L.C. Gallo,$^{9}$
M. Berton,$^{10}$
R. Blandford,$^{11}$
J.L. G\'omez,$^{12}$
and A.G. Gonzalez$^{9}$
\\
$^{1}$ Max-Planck-Institut f\"ur Radioastronomie, 
Auf dem H{\"u}gel 69, 53121 Bonn, Germany\\
$^{2}$ Department of Physics, Geology, and Engineering Technology, Northern Kentucky University, 1 Nunn Dr, Highland Heights, KY 41099, USA\\
$^{3}${Center for Astrophysics $\vert{}$ Harvard~ \& ~Smithsonian, Cambridge, MA~02138, USA}\\
$^{4}$ Department of Astronomy, Columbia University, New York, NY 10027, USA \\
$^{5}$ Department of Physics, Columbia University, New York, NY 10027, USA \\
$^{6}$ Department of Astronomy, School of Physics, Peking University, Beijing 100871, People’s Republic of China\\
$^{7}$ Kavli Institute for Astronomy and Astrophysics, Peking University, Beijing 100871, People’s Republic of China\\
$^{8}$ Department of Physics \& Astronomy, Northwestern University, Evanston, IL 60208, USA \\
$^{9}$ Department of Astronomy and Physics, Saint Mary’s University, 923 Robie Street, Halifax, NS, B3H 3C3, Canada\\
$^{10}$ European Southern Observatory (ESO), Alonso de Córdova 3107, Casilla 19, Santiago 19001, Chile\\
$^{11}$ Kavli Institute for Particle Astrophysics and Cosmology (KIPAC), Stanford University, Stanford, CA 94305, USA\\ 
$^{12}$ Inst. de Astrofísica de Andalucía-CSIC, Glorieta de la Astronomía s/n, E-18008 Granada, Spain\\
}

\date{Accepted 2023 February 7. Received 2023 January 27; in original form 2022 December 22}

\pubyear{2023}

\begin{document}
\label{firstpage}
\pagerange{\pageref{firstpage}--\pageref{lastpage}}
\maketitle

%% word constraint:  max 300 words for main paper, max 200 for letter. 
\begin{abstract}
%% note: abstract slighly lger 
The project MOMO (Multiwavelength Observations and Modelling of OJ\,287) was set up to test predictions of binary supermassive black hole (SMBH) scenarios and to understand disk-jet physics of the blazar OJ\,287. 
After a correction, the precessing binary (PB) SMBH model predicted the next main outburst of OJ\,287 in 2022 October, no longer in July, 
making the outburst well observable and the model testable. 
We have densely covered this period in our ongoing multi-frequency radio, optical, UV, and X-ray monitoring.
The predicted outburst was not detected.
 Instead, OJ\,287 was at low optical--UV emission levels, declining further into November.  The predicted thermal bremsstrahlung spectrum was not observed either, at any epoch.
Further, applying scaling relations, we estimate a SMBH mass of OJ\,287 of 10$^8$ M$_{\odot}$.  
The latest in a sequence of deep low-states that recur every 1--2 yrs
is used to determine an upper limit on  
the Eddington ratio and on the accretion-disk luminosity. This limit is at least a factor of 10 lower than required by the PB model 
with its massive primary SMBH of $>10^{10}$ $M_{\odot}$. 
All these results favor 
alternative binary SMBH models of OJ\,287 that neither require strong orbital precession nor a very large mass of the primary SMBH. 
\end{abstract}


%%%%%%%%%%%%%%%%%%%%%%%%%%%%%%%%%%%%%%%%%%%%%%%%%%

%%%%%%%%%%%%%%%%% BODY OF PAPER %%%%%%%%%%%%%%%%%%

\section{Introduction}

Binary  SMBHs play an important role in extragalactic astrophysics. They are an important ingredient in our understanding of galaxy evolution and SMBH demographics, 
and during their final coalescence they are the loudest known sources of gravitational waves in the Universe 
\citep[e.g.][]{Begelman1980, Volonteri2003, Komossa2016}. 
They reveal their presence by semi-periodicity in light curves or in spatial structures.
Other mechanisms of periodicity that do not necessarily require a binary have been discussed as well \citep{Liska2018}. 
In cases where only very few periods have been observed, red noise is a possible alternative explanation \citep{Uttley2005}. 
Therefore each case requires a multiwavelength approach, long-term monitoring, and careful modelling \citep[e.g.][]{Graham2015, DOrazio2015, Dotti2023}.
One of the best-observed binary SMBH candidates is the blazar OJ 287 at redshift $z$=0.306. Its claim to fame is its
bright optical outbursts that repeat every 11--12 years \citep{Sillanpaa1988}, most recently in 2016--2017 \citep{Komossa2023-apj}. Each of these outbursts exhibits a double-peak structure, with peak-to-peak separation of order one year \citep{Sillanpaa1996}.
Different binary SMBH scenarios have been suggested to explain these observations, whereas \citet{Villforth2010} preferred a non-binary interpretation.
One class of binary models assumes that the outbursts represent true changes in the luminosity, driven either by accretion-disk impacts of the secondary SMBH and/or by increases in accretion rate on the primary after periastron passage of the secondary SMBH
\citep{Lehto1996, Valtaoja2000, Liu2002}. The other class of binary models assumes that outbursts are the consequence of relativistic beaming in one or two jetted SMBHs \citep{Katz1997, Villata1998, Britzen2018}. 

The best explored binary scenario is based on detailed orbital modelling of the system. It requires a high primary mass of 1.8 $\times$ 10$^{10}$ M$_{\odot}$ and a secondary SMBH in a highly eccentric, highly inclined, and highly precessing orbit that impacts the primary's accretion disk twice during each orbit \citep[precessing binary (PB) model hereafter;][]{Lehto1996, Valtonen2021}. 
The PB model claims very high accuracy in predicting the 
timing of future outbursts, on the order of one to a few days, despite the 
complex physics involved, 
and for events that are separated by more than a decade. 

\begin{figure*}
\centering
\includegraphics[clip, trim=0.9cm 5.3cm 1.0cm 2.3cm, width=13.9cm]{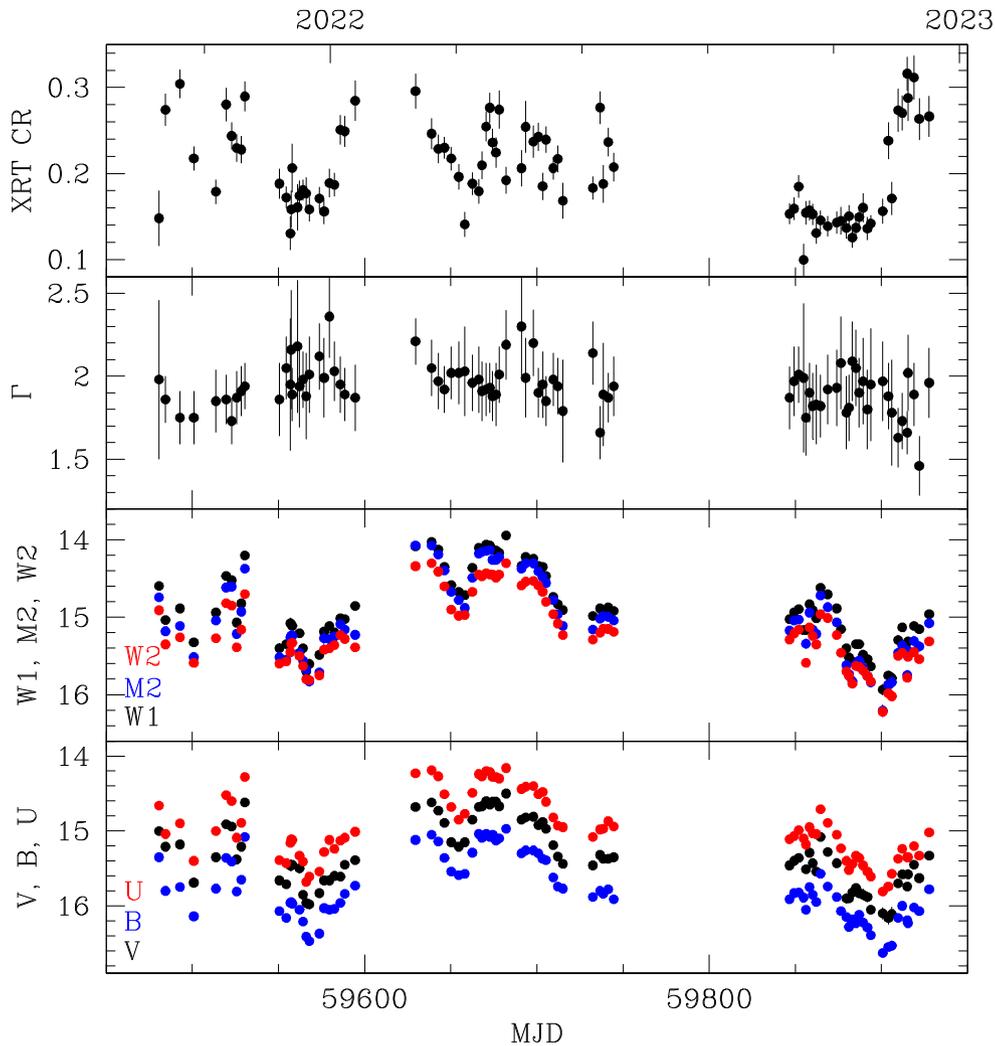}
    \caption{Swift light curve of OJ 287 between 2021 September and 2022 December 15. From top to bottom: observed Swift XRT (0.3--10 keV) count rate, X-ray photon index,  and observed UVOT UV and optical magnitudes in the VEGA system not corrected for Galactic extinction. 
    The gap in Swift observations between mid-June and mid-September is due to the proximity of OJ 287 to the Sun, such that it becomes unobservable with Swift each year. Never does the light curve reach the bright emission levels of the double-peaked main outbursts of OJ 287 (with V magnitudes then exceeding 13-12th mag). 
    In late 2022, low-amplitude variability is ongoing that is not different from that at any other epoch. The predicted 2022 October main outburst was not observed.  
   }
    \label{fig:MWL-Swift}
\end{figure*}

If this precision can be achieved, then OJ 287 is of great interest for high-precision tests of general relativity and gas physics, for instance, (1) binary orbital shrinking due to gravitational wave emission \citep{Centrella2010}, (2) the black hole no-hair theorem \citep{Valtonen2008, Dey2018}, (3) the presence of a dark matter spike surrounding the central SMBH \citep{Alachkar2022}, and/or (4) the microphysics of potential mechanisms producing the outbursts \citep{Ivanov1998}.
Therefore, it is of great interest to test  
predictions of the PB model.
These tests are one of the two main goals of the project MOMO \citep[Multiwavelength Observations and Modelling of OJ 287;][]{Komossa2021-universe}, which was set up in 2015 (the other goal is understanding blazar disk-jet physics). In the course of MOMO, dense, long-term, multifrequency monitoring from radio to X-rays is carried out, along with dedicated deeper follow-up spectroscopy and imaging with various ground- and space-based facilities when OJ 287 experiences outbursts, deep fades, and/or extreme spectral states \citep[e.g.][]{Komossa2017, Komossa2020, Komossa2021-apj, Komossa2022-mnras}. 

In the framework of the PB model, the latest prediction was that the latest of the bright main outbursts of OJ 287 should occur in October 2022, on October 10 $\pm{10}$ days \citep{Valtonen2022}.  This date supersedes the earlier prediction for July 2022, after 
a correction that amounts to $\sim$3 months. While in July 2022, OJ 287 would have been unobservable with most observatories due to its close proximity to the Sun, it became re-observable in September and was therefore covered in our ongoing monitoring program.
Here, we present dense multiwavelength coverage of this period, we report the absence of the predicted main outburst, 
we derive the SMBH mass and Eddington ratio of OJ 287, 
and we discuss implications for the different binary SMBH scenarios of OJ 287.

Throughout this Letter, timescales and frequencies are given in the observer's frame when reporting measurement results, unless noted otherwise. 
We use a cosmology with 
$H_{\rm 0}$=70 km\,s$^{-1}$\,Mpc$^{-1}$, $\Omega_{\rm M}$=0.3 and $\Omega_{\rm \Lambda}$=0.7. At the distance of OJ 287, this corresponds to a scale of 4.5 kpc/arcsec \citep{Wright2006}. 

\section{Data acquisition and analysis}

Our observations have been carried out with the Neil Gehrels Swift observatory \citep[Swift hereafter;][]{Gehrels2004} with the UV--optical telescope (UVOT) in three optical and three UV bands, and with the X-ray telescope (XRT) in the (0.3-10 keV) X-ray regime, and with the Effelsberg 100\,m radio telescope \citep{Kraus2003} at frequencies between 2.6 and 44 GHz.  
The Swift monitoring cadence is 2--3 days. The Effelsberg cadence is about three weeks and also depends on weather conditions. 

\begin{figure}
\centering
\includegraphics[clip, trim=1.6cm 5.4cm 1.0cm 2.8cm, width=\columnwidth]{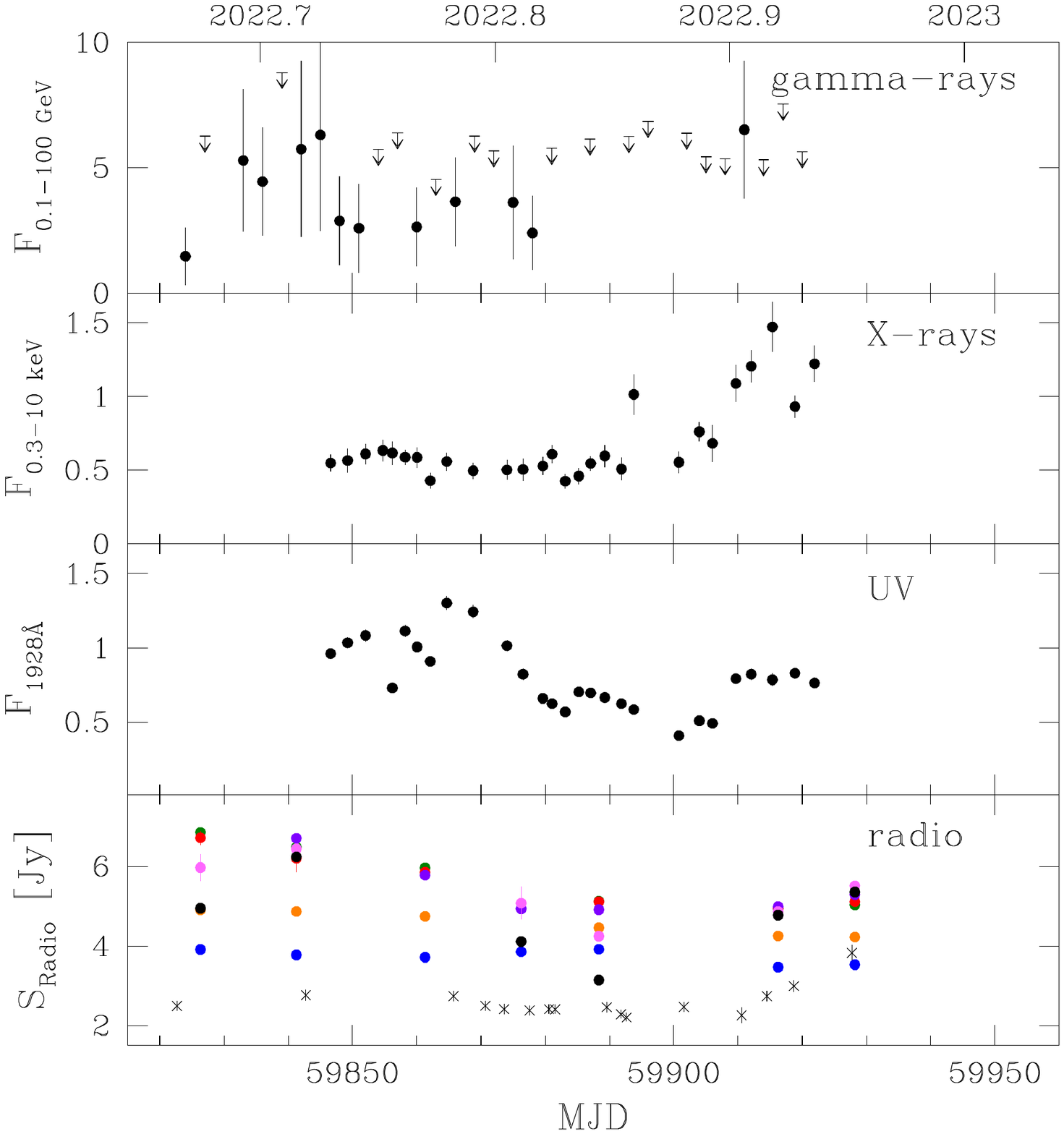}
    \caption{Multiwavelength (MWL) light curve of OJ 287 between 2022 September and 2022 December 15. From top to bottom: Fermi $\gamma$-rays (3 day average), Swift X-rays, Swift UV (at $\lambda_{\rm obs}$=1928\AA), Effelsberg and SMA radio observations at selected frequencies (2.6 GHz: blue, 4.85 GHz: orange, 10.45 GHz: green, 14.25 GHz: red, 19.25 GHz: purple, 36.25 GHz: pink, 43.75 GHz: black, SMA data at 1.3 mm: black crosses).  
    The $\gamma$-ray flux (0.1--100 GeV band; one-week averages), the absorption-corrected X-ray flux (0.3--10 keV band), and the extinction-corrected observed UV flux at $\lambda_{\rm obs}$=1928\AA~are given in units of 10$^{-11}$ erg s$^{-1}$ cm$^{-2}$. The radio flux densities are given in Jansky. }
    \label{fig:MWL-all}
\end{figure}

Data reduction was performed in a standard manner that is described in detail by \citet[][Swift]{Komossa2021-apj} and \citet[][Effelsberg]{Komossa2022-mnras}. 
The Swift light curve is shown in Fig. \ref{fig:MWL-Swift}. 
Public Fermi data  \citep{Kocevski2021} in the way included before as described by \citet{Komossa2021-apj}, and public Submillimeter Array (SMA) data \citep{Gurwell2007} were added to our analysis and are shown in Fig. \ref{fig:MWL-all} together with our multifrequency Effelsberg data. 

\section{Results}
\label{sec:results} 

\subsection{Absence of predicted 2022 October outburst} 
During 2022 October, and more generally in the time interval September--December, 
only small amplitude flaring, if any, was detected, of the kind that is always seen in OJ 287 \citep[cf][for the longterm light curve] {Komossa2021-apj}. None stands out in amplitude (Fig. \ref{fig:MWL-Swift}, \ref{fig:MWL-all}). In late October, the emission dropped by 1.2 magnitudes 
into a deep low-state and started to rise again from November 20 on. Emission levels remain low in December. 
The predicted outburst in October was not observed. 

\subsection{Absence of predicted optical thermal bremsstrahlung spectrum}

\begin{figure}
\centering
\includegraphics[clip, angle=-90, trim=1.0cm 1.3cm 2.4cm 2.9cm, width=\columnwidth]{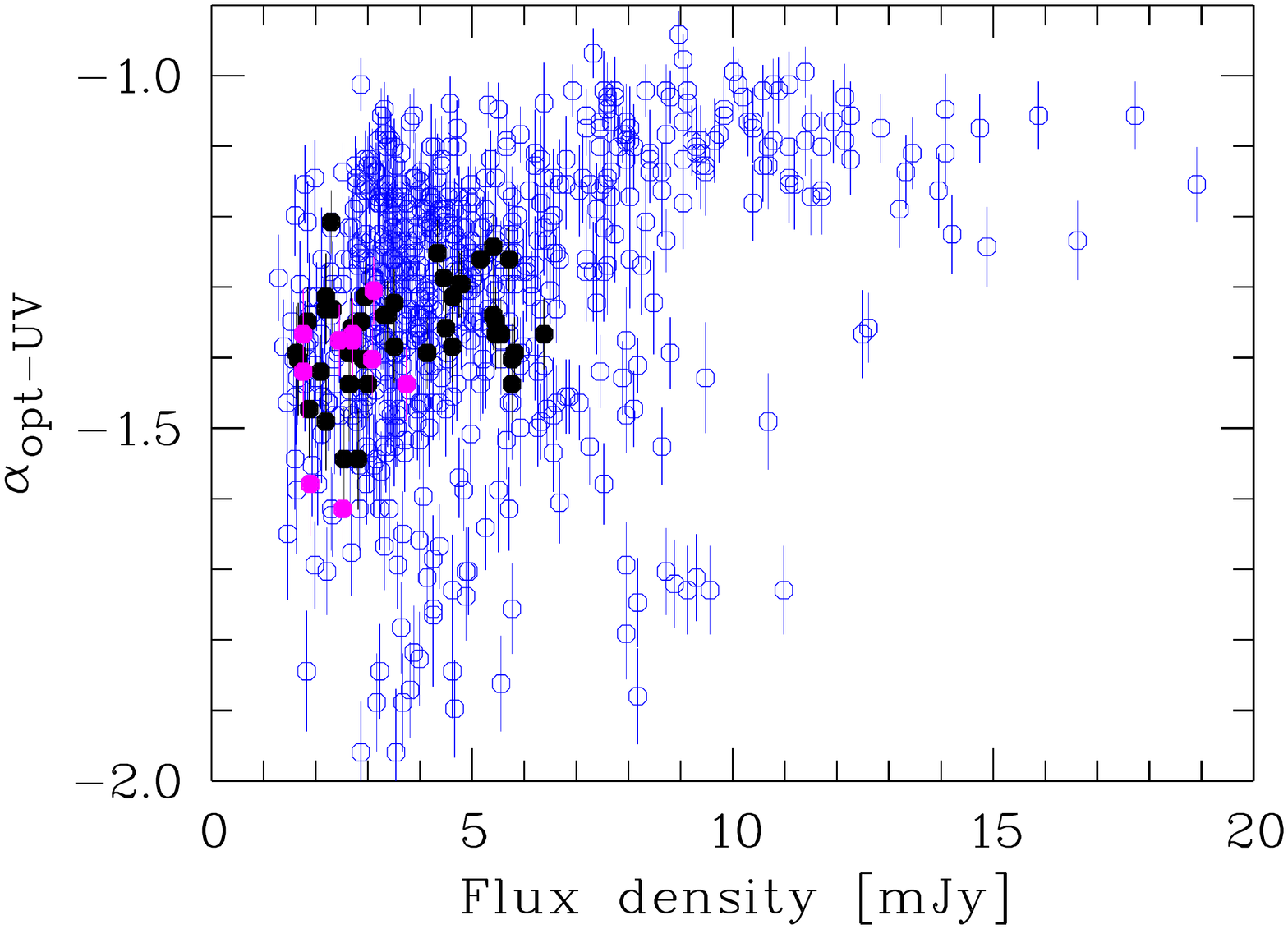}
\caption{Spectral index $\alpha_{\rm \nu, opt-UV}$ versus the V-band flux density at $\lambda_{\rm obs}$=5468\AA~ 
between 2005 and 2022 December15. The time period 2021 December -- 2022 June is marked in black, and the period of 2022 October is marked in magenta. A spectral index of --0.2, required by the PB model, is not observed at any time. 
}
    \label{fig:spectral-index}
\end{figure}

Next, we might speculate that we missed the October outburst because it did happen, but escaped detection because of an unfortunate, never-before-seen fading of the intrinsic blazar emission at the very same epoch, such that the total emission did not increase much. The small flaring we see in September or October would then represent a much larger intrinsic flare. However, we should then still clearly see the optical--UV thermal bremsstrahlung spectrum with a spectral index $\alpha_{\nu}$=--0.2, a main prediction of the PB model \citep[e.g., Sect. 3 of][]{Valtonen2021, Valtonen2022}, where 
the observed spectral index $\alpha_{\nu}$ is defined as 
$S_\nu \propto \nu^{+\alpha_\nu}$. 
Instead, we find that $\alpha_{\nu, \rm {opt-UV}}$ varies between --1.3 and --1.6 (or $\alpha_{\nu, \rm {opt}}$ between --1 and --1.35 using V and U). These values are the same as those seen at other epochs, and they never reach the predicted $\alpha_{\nu}$=--0.2 of thermal bremsstrahlung (Fig. \ref{fig:spectral-index}). 
The average spectral index of all Swift observations of OJ 287 between 2005 and 2022 was $\alpha_{\nu, \rm opt-UV}$=--1.30$\pm{0.18}$.

\subsection{MWL variability}
The X-ray emission has been rather constant since 2022 September. It only started to rise in late November increasing by a factor 2 in countrate.  

In the Fermi $\gamma$-ray band, OJ 287 remains rather inactive; similar to recent years but unlike its epochs of bright flaring observed prior to 2016 \citep{Abdo2009}.  

The radio emission shows an overall decline after the non-thermal flare \citep{Komossa2022-mnras} of 2021 November--2022 June and reaches a minimum in 2022 November (see next Section). At high frequencies, a small flare is detected in 2022 September, and a low-state is reached in November--December, followed by a sharp rise in mid-December. 

\subsection{Deep fade}
A new UV--optical deep fade with its minimum on 2022 November 17--20 is detected. 
While it was speculated that a previous optical--radio deep fade in 1989 was perhaps related to jet deflection of the primary SMBH during a close passage of the secondary SMBH in the context of binary models of OJ 287 \citep{Takalo1990}, the timing of the deep fade of 2017 was inconsistent with such an interpretation, given predictions of the PB model for the location of the secondary SMBH at that epoch \citep{Komossa2021-apj}.  
Since 2017, the deep fades repeated, ruling out such an interpretation in any case. 
The 2022 deep fade is the latest in a sequence of similar such events \citep{Komossa2022-mnras}.  It is accompanied by a minimum in radio emission as well, well covered with the SMA. At 1.3 mm, values as low as 2 mJy are reached, comparable to previous low states in 2015 and 2017. 
The observations lend further support to findings of a possible repetition of these events on a timescale of 1--2 years \citep{Komossa2023-apj}, consistent with earlier reports of 1--2 year periodicities in optical and radio light curves of OJ 287 \citep[e.g.][]{Hughes1998, Bhatta2016, Britzen2018}. 
These may reflect the timescale of shocks in the jet \citep{Hughes1998}, 
jet rotation \citep{Britzen2018}, jet production efficiency 
\citep{Bhatta2016}, or could still be indirectly related to binary disk-crossing events if these trigger resonances in the disk, or disk oscillations \citep[][]{Liu2006}, even though binary disk crossings are already invoked to explain the $\sim$11 yr period of OJ 287. 
Independent of the underlying physical mechanism, the 2022 November deep fade is used below to estimate an upper limit on the accretion disk luminosity and the Eddington ratio.    

\section{Discussion and conclusions} 

Several binary SMBH scenarios have been suggested for OJ 287 (Sect. 1). The majority of them does not make high-precision predictions for future outbursts except that these should be semi-periodic with an uncertainty of order a year, based on previous observations. The PB model has modelled OJ 287 in detail. It explains the lack of strict periodicity of previous outbursts by invoking a very massive primary SMBH and consequently a highly precessing orbit of the secondary SMBH. In that model, due to the high forward precession that shifts the timing of disk impacts, main bursts and their two peaks no longer occur at semi-periodic intervals.
However, the predicted 2022 outburst was not observed, as shown above. 

Next, we discuss other tests of the PB model, in comparison with other binary scenarios. For instance, in order to work, the PB model requires a mass of the primary SMBH larger than 10$^{10}$ M$_{\odot}$ and a very high Eddington ratio $L_{\rm disk}/L_{\rm Edd}$=0.08 \citep[e.g., Sect 2 of][]{Valtonen2021}.   

The high primary SMBH mass is not implausible {\sl{per se}}, as other SMBHs at that mass range have been found in large spectroscopic data bases such as the Sloan Digital Sky Survey \citep{Shen2011}. 
However, no massive host galaxy of OJ 287 has been detected so far in optical--NIR imaging \citep[][and references therein]{Nilsson2020}, leaving only a small chance for an extended, so far undetected but massive halo.
The measured host magnitude places 
OJ 287 in the middle of the host galaxy luminosity distribution of BL Lac objects. 

Next, we use broad-line region (BLR) scaling relations to estimate the black hole mass of OJ 287, based
on the virial relation established for broad-line AGN, using the line
width of broad emission lines. Since the optical continuum luminosity is dominated by the jet emission, we use the 
line luminosity instead. The SMBH mass is given by \citep{Vestergaard2006}   
\begin{equation}
M_{\rm BH} = 10^{6.67}~~ (\frac{L_{\rm H\beta}} {10^{42}\, \rm{erg\,s}^{-1}})^{0.63} ~~(\frac{\rm{FWHM(H\beta)}} {1000\, \rm{km\,s}^{-1}})^{2} ~ M_{\odot}. 
\end{equation}
Since only H$\alpha$ is well measured in OJ 287, we 
use its width and a recombination value of 3.1 for the intensity ratio of the two Balmer lines \citep{Osterbrock1989}.  
With FWHM(H$\alpha$)=4200 km\,s$^{-1}$ 
and a high-state luminosity of $L_{\rm H\alpha}$=10$^{42.8}$ erg\,s$^{-1}$=3.1\,$L_{\rm H\beta}$ \citep{Sitko1985, Nilsson2010}, 
we obtain a SMBH mass of OJ 287 of $M_{\rm BH}=1.3 \times 10^{8}$ M$_{\odot}$. 

Next, we show that the high Eddington ratio, $L_{\rm disk}/L_{\rm Edd}$, required by the PB model is not confirmed by observations, since the model overpredicts the observed luminosity by at least a factor of 10.    
First, from XMM-Newton observations, \citet{Komossa2021-mnras} derived a tight constraint on $L_{\rm X}/L_{\rm Edd}<5.6 \times 10^{-4}$ (using $M_{\rm BH, primary}$=1.8\,10$^{10}$ M$_{\odot}$ from the PB model);  
too small for a such a massive SMBH that should have a detectable accretion-disk corona  
even if the disk emission itself did not reach into the X-ray band. 
Further, at typical bolometric correction factors, the upper limit on $L_{\rm bol}/L_{\rm Edd}$ from observations is much smaller than the value required by the PB model \citep{Komossa2021-mnras}.  
Here, we extend that estimate using the optical observations during the 
deep low-state in 2022 to obtain an upper limit on any disk emission. 
The low-state is still dominated by the jet, so any disk emission must be much fainter than that.  
In low-state, $L_{\rm V}$=2.3 $\times$ 10$^{45}$ erg s$^{-1}$. Then, $L_{\rm disk}$ $\ll$ $L_{\rm bol}$=2.1 $\times$ 10$^{46}$ erg s$^{-1}$ using a bolometric correction factor of 9 \citep{Kaspi2000}.  
Therefore, $L_{\rm disk}/L_{\rm Edd}$ $\ll$ 0.009 for the initial PB model SMBH mass of $M_{\rm BH}$=1.8 $\times$ 10$^{10}$ M$_{\odot}$, at least a factor of 10 lower than  required by the PB model. 

We can also directly estimate the disk luminosity from the observed H$\alpha$ luminosity, in which case we find $L_{\rm disk, H\alpha}$ = 1.75 $\times$ 10$^{45}$ erg s$^{-1}$, using \citet{Zhou2006} and the same bolometric correction as before \citep{Kaspi2000}. This is a further factor of 12 lower than the upper limit estimated above from the 2022 November low-state, and then gives $L_{\rm disk, H\alpha}/L_{\rm Edd}$ = 7 $\times$ 10$^{-4}$ when using the PB model value of $M_{\rm BH}$=1.8 $\times$ 10$^{10}$ M$_{\odot}$. 

In summary, the required $L_{\rm disk}/L_{\rm Edd}\sim 0.1$ in the PB model strongly overpredicts the actually observed value. If instead, we use the observed Eddington ratio of $L_{\rm disk, H\alpha}/L_{\rm Edd}\ll 0.01$ that would be implied for a massive primary SMBH exceeding 10$^{10}$ M$_{\odot}$, then the accretion disk is expected to be in the (geometrically thick, optically thin) advection-dominated accretion flow (ADAF) mode, no longer in the standard disk mode as required by the PB model. In the ADAF mode, the 
mechanism of producing sharp impact flares no longer works either \citep{Villforth2010}. 
At a more moderate (primary) black hole mass of order $M_{\rm BH}$=10$^{8}$ M$_{\odot}$ that we directly estimated from BLR scaling relations here, one obtains $L_{\rm disk, H\alpha}/L_{\rm Edd}$=0.13, near the upper end of, and consistent with, Eddington ratios observed in BL Lac objects \citep[e.g.][]{Foschini2015}. These results strongly favor alternative OJ 287 models that are based on lower masses of the primary SMBH \citep{Valtaoja2000, Britzen2018, Liu2002}. 

Finally, we note that the 2021 precursor flare predicted by the PB model was not observed either. It was predicted to happen in 2021 December \citep{Valtonen2021}, but was found to be absent \citep{Komossa2022-an} not only in December but also in subsequent months.

We conclude that the new observations and new parameter estimates of OJ 287 do not support the parameter values and predictions of the PB model.
The model requires a very massive primary SMBH, a high Eddington ratio of a standard geometrically thin accretion disk,  and a large disk luminosity. In contrast, we derive a (primary) SMBH mass that is a factor 100 lower, and measure a disk luminosity that is at least a factor 10-100 lower than required. Furthermore, the main outburst and the precursor flare predicted by the model (to happen in 12/2021, and 10/2022, respectively) were not detected either. Neither was the predicted thermal Bremsstrahlung spectrum observed with Swift at any time in 2021--2022. 
Our results favor binary models of OJ 287 (Sect. 1) that are based on a much less massive primary SMBH. 
Revisions of these models are still needed to account for  
the fact that the main double-peaked outbursts are not strictly periodic but appear with a scatter of $\pm{1}$ yr. The last two main outbursts came about a year early,  in 2005--2006 \citep{Villforth2010} and in 2016--2017 \citep{Komossa2023-apj}. 

Monitoring of OJ 287 continues in the course of the MOMO project, with the long-term goal of covering 1--2 orbital periods. 

\section*{Acknowledgements}

It is our pleasure to thank the Swift team for carrying out the observations we proposed.
This work made use of data supplied by the UK Swift Science Data Centre at the University of Leicester \citep{Evans2007}. 
This work is partly based on data obtained with the 100\,m telescope of the Max-Planck-Institut f\"ur Radioastronomie at Effelsberg.   
The Submillimeter Array near the summit of Maunakea is a joint project between the Smithsonian Astrophysical 
Observatory and the Academia Sinica Institute of Astronomy and Astrophysics and is funded by the Smithsonian 
Institution and the Academia Sinica. 
The authors recognize and acknowledge the very significant cultural role 
and reverence that the summit
of Maunakea has always had within the indigenous Hawaiian community. We are most 
fortunate to have the opportunity to conduct
observations from this mountain.
Data from this program are available at the 
SMA website {\url{http://sma1.sma.hawaii.edu/callist/callist.html}}. 
This work has made use of Fermi-LAT data supplied by
\citet{Kocevski2021} at \url{https://fermi.gsfc.nasa.gov/ssc/data/access/lat/LightCurveRepository/}. Z.H. acknowledges NASA ATP grant 80NSSC22K082.  
This research was supported in part by the National Science Foundation under Grant No. NSF PHY-1748958. 

%%%%%%%%%%%%%%%%%%%%%%%%%%%%%%%%%%%%%%%%%%%%%%%%%%
\section*{Data Availability}
Our Effelsberg radio data and reduced Swift data are available on reasonable request. The raw Swift data of our project are available in the Swift archive at \url{https://swift.gsfc.nasa.gov/archive/}. 



\bibliographystyle{mnras}
% \bibliography{references} % if your bibtex file is called references.bib

\begin{thebibliography}{99}

\bibitem[\protect\citeauthoryear{Abdo et al.}{2009}]{Abdo2009}
Abdo A.A., et al. 2009, ApJ, 700, 597

\bibitem[\protect\citeauthoryear{Alachkar et al.}{2022}]{Alachkar2022}
Alachkar A., Ellis J., Fairbairn M., 2022, arXiv e-prints, arXiv:2207.10021

\bibitem[\protect\citeauthoryear{Begelman et al.}{1980}]{Begelman1980}
Begelman M.C., Blandford R.D., Rees M.J. 1980, Nature, 287, 307

\bibitem[\protect\citeauthoryear{Bhatta et al.}{2016}]{Bhatta2016}
Bhatta G., et al., 2016, ApJ, 832, 47

\bibitem[\protect\citeauthoryear{Britzen et al.}{2018}]{Britzen2018}
Britzen S., et al. 2018, MNRAS, 478, 3199

\bibitem[\protect\citeauthoryear{Centrella et al.}{2010}]{Centrella2010}
Centrella J., Baker J. G., Kelly B. J., van Meter J. R., 2010, Reviews of
Modern Physics, 82, 3069

\bibitem[\protect\citeauthoryear{D'Orazio et al.}{2015}]{DOrazio2015}
D’Orazio D. J., Haiman Z., Schiminovich D., 2015, Nature, 525, 351

\bibitem[\protect\citeauthoryear{Dey et al.}{2018}]{Dey2018} 
Dey L., et al. 2018, ApJ, 866, 11

\bibitem[\protect\citeauthoryear{Dotti et al.}{2023}]{Dotti2023} 
Dotti M., et al., 2023, MNRAS, 518, 4172

\bibitem[\protect\citeauthoryear{Evans et al.}{2007}]{Evans2007} 
Evans P. A., et al., 2007, A\&A, 469, 379

\bibitem[\protect\citeauthoryear{Foschini et al.}{2015}]{Foschini2015} 
Foschini L., et al., 2015, A\&A, 575, A13

\bibitem[\protect\citeauthoryear{Gehrels et al.}{2004}]{Gehrels2004} 
Gehrels N., et al. 2004, ApJ 611, 1005

\bibitem[\protect\citeauthoryear{Graham et al}{2015}]{Graham2015} 
Graham M. J., et al., 2015, MNRAS, 453, 1562

\bibitem[\protect\citeauthoryear{Gurwell et al.}{2007}]{Gurwell2007}
Gurwell M.A., Peck A.B., Hostler S.R., Darrah, M.R., Katz C.A., 2007, 
in From Z-Machines to ALMA: (Sub)Millimeter Spectroscopy of Galaxies, ASPC 375, 234

\bibitem[\protect\citeauthoryear{Hughes et al.}{1998}]{Hughes1998}
Hughes P. A., Aller H. D., Aller M. F., 1998, ApJ, 503, 662

\bibitem[\protect\citeauthoryear{Ivanov et al.}{1998}]{Ivanov1998}
Ivanov P.B., Igumenshchev I.V., Novikov I.D., 1998, ApJ 507, 131

\bibitem[\protect\citeauthoryear{Kaspi}{2000}]{Kaspi2000}
Kaspi S., Smith P. S., Netzer H., Maoz D., Jannuzi B. T., Giveon U., 2000, ApJ, 533, 631 

\bibitem[\protect\citeauthoryear{Katz}{1997}]{Katz1997}
Katz J.I., 1997, ApJ 478, 527

\bibitem[\protect\citeauthoryear{Kocevski}{2021}]{Kocevski2021}
Kocevski D., Valverde J., Garrappa S., Negro M., Brill A., Ballet J., Lott B.,
2021, The Astronomer’s Telegram, 15110, 1

\bibitem[\protect\citeauthoryear{Komossa et al.}{2016}]{Komossa2016}
Komossa S., Baker J. G., Liu F. K., 2016, IAU Focus Meeting, 29B, 292

\bibitem[\protect\citeauthoryear{Komossa et al.}{2017}]{Komossa2017}
Komossa S., et al. 2017, IAUS 324, 168

\bibitem[\protect\citeauthoryear{Komossa et al.}{2020}]{Komossa2020}
Komossa S., Grupe D., Parker M. L.,Valtonen M. J., Gómez J. L., Gopakumar A., Dey L., 2020, MNRAS, 498, L35

\bibitem[\protect\citeauthoryear{Komossa et al.}{2021a}]{Komossa2021-universe}
Komossa S., et al., 2021a, Universe, 7, 261

\bibitem[\protect\citeauthoryear{Komossa et al.}{2021b}]{Komossa2021-mnras}
Komossa S., et al., 2021b, MNRAS, 504, 5575

\bibitem[\protect\citeauthoryear{Komossa et al.}{2021c}]{Komossa2021-apj}
Komossa S., Grupe D., Gallo L. C., Gonzalez A., Yao S., Hollett A. R., Parker M. L., Ciprini S., 2021c, ApJ, 923, 51

\bibitem[\protect\citeauthoryear{Komossa et al.}{2022a}]{Komossa2022-an}
Komossa S., et al., 2022a, AN, in press, arXiv e-prints, arXiv:2207.11291

\bibitem[\protect\citeauthoryear{Komossa et al.}{2022b}]{Komossa2022-mnras}
Komossa S., et al., 2022b, MNRAS, 513, 3165

\bibitem[\protect\citeauthoryear{Komossa et al.}{2023}]{Komossa2023-apj}
Komossa S., Kraus A., Grupe D., Gonzalez A. G., Gurwell M. A., Gallo L. C., Liu F. K., et al. 2023, ApJ, in press

\bibitem[\protect\citeauthoryear{Kraus et al.}{2003}]{Kraus2003}
Kraus A., et al. 2003, A\&A, 401, 161

\bibitem[\protect\citeauthoryear{Lehto \& Valtonen}{1996}]{Lehto1996}
Lehto H.J., Valtonen M.J., 1996, ApJ, 460, 207

\bibitem[\protect\citeauthoryear{Liska et al.}{2018}]{Liska2018} Liska M., Hesp C., Tchekhovskoy A., Ingram A., van der Klis M., Markoff
S., 2018, MNRAS, 474, L81

\bibitem[\protect\citeauthoryear{Liu \& Wu}{2002}]{Liu2002}
Liu F. K., Wu X. B., 2002, A\&A, 388, L48

\bibitem[\protect\citeauthoryear{Liu et al.}{2006}]{Liu2006}
Liu F. K., Zhao G., Wu X.-B., 2006, ApJ, 650, 749

\bibitem[\protect\citeauthoryear{Nilsson et al.}{2010}]{Nilsson2010} Nilsson K., Takalo L. O., Lehto H. J., Sillanpää A., 2010, A\&A, 516, A60

\bibitem[\protect\citeauthoryear{Nilsson et al.}{2020}]{Nilsson2020} Nilsson K., et al., 2020, ApJ, 904, 102

\bibitem[\protect\citeauthoryear{Osterbrock}{1989}]{Osterbrock1989} Osterbrock D. E., 1989, Astrophysics of gaseous nebulae and active galactic
nuclei

\bibitem[\protect\citeauthoryear{Shen}{2011}]{Shen2011} Shen Y., et al., 2011, ApJS, 194, 45

\bibitem[\protect\citeauthoryear{Sillanp\"a\"a et al.}{1988}]{Sillanpaa1988} Sillanpaa A., Haarala S., Valtonen M. J., Sundelius B., Byrd G. G., 1988,
ApJ, 325, 628



\bibitem[\protect\citeauthoryear{Sillanp\"a\"a et al.}{1996}]{Sillanpaa1996}
Sillanp\"a\"a A., et al. 1996, A\&A, 315, L13

\bibitem[\protect\citeauthoryear{Sitko}{1985}]{Sitko1985}Sitko M. L., Junkkarinen V. T., 1985, PASP, 97, 1158

\bibitem[\protect\citeauthoryear{Takalo}{1990}]{Takalo1990}Takalo L. O., Kidger M., de Diego J. A., Sillanpaa A., Piirola V., Terasranta
H., 1990, A\&AS, 83, 459

\bibitem[\protect\citeauthoryear{Uttley}{2005}]{Uttley2005}Uttley P., McHardy I. M., Vaughan S., 2005, MNRAS, 359, 345

\bibitem[\protect\citeauthoryear{Valtaoja et al.}{2000}]{Valtaoja2000}
Valtaoja E., Teräsranta H., Tornikoski M., Sillanpää A., Aller M. F., Aller
H. D., Hughes P. A., 2000, ApJ, 531, 744

\bibitem[\protect\citeauthoryear{Valtonen et al.}{2008}]{Valtonen2008}Valtonen M. J., et al., 2008, Nature, 452, 851

\bibitem[\protect\citeauthoryear{Valtonen et al.}{2021}]{Valtonen2021}
Valtonen M.J., et al. 2021, Galaxies, 10, 1 

\bibitem[\protect\citeauthoryear{Valtonen et al.}{2022}]{Valtonen2022}Valtonen M. J., et al., 2022, arXiv e-prints, arXiv:2209.08360, version 1

\bibitem[\protect\citeauthoryear{Vestergaard et al.}{2006}]{Vestergaard2006}Vestergaard M., Peterson B. M., 2006, ApJ, 641, 689

\bibitem[\protect\citeauthoryear{Villata et al.}{1998}]{Villata1998}Villata M., Raiteri C. M., Sillanpaa A., Takalo L. O., 1998, MNRAS, 293, L13

\bibitem[\protect\citeauthoryear{Villforth et al.}{2010}]{Villforth2010}
Villforth C., et al. 2010, MNRAS, 402, 2087

\bibitem[\protect\citeauthoryear{Volonteri et al.}{2003}]{Volonteri2003}
Volonteri M.,  Haardt F., Madau P, 2003, ApJ, 582, 559

\bibitem[\protect\citeauthoryear{Wright}{2006}]{Wright2006}
Wright E.L., 2006, PASP, 118, 1711

\bibitem[\protect\citeauthoryear{Zhou et al.}{2006}]{Zhou2006}Zhou H., Wang T., Yuan W., Lu H., Dong X., Wang J., Lu Y., 2006, ApJS,
166, 128


%\begin{thebibliography}{99}
\end{thebibliography}

%%%%%%%%%%%%%%%%%%%%%%%%%%%%%%%%%%%%%%%%%%%%%%%%%%

%%%%%%%%%%%%%%%%% APPENDICES %%%%%%%%%%%%%%%%%%%%%

% \appendix

% \section{Some extra material}

%%%%%%%%%%%%%%%%%%%%%%%%%%%%%%%%%%%%%%%%%%%%%%%%%%

% Don't change these lines
\bsp	% typesetting comment
\label{lastpage}
\end{document}